\documentclass[journal]{IEEEtran}

\ifCLASSINFOpdf

\else

\fi

\usepackage{multirow}
\usepackage{float}
\usepackage{hyperref}
\usepackage[export]{adjustbox}
\usepackage{xcolor}
\usepackage{caption}
\usepackage{subcaption}
\usepackage{afterpage}
\usepackage[labelsep=period, figurename=Figure]{caption}

\usepackage{amsmath}
\usepackage{verbatim}
\usepackage{graphicx}

\usepackage{tikz}
\usepackage{textcomp}
\usepackage{hyperref}
\usepackage{lipsum}

\newcommand\copyrighttext{%
  \footnotesize This article has been accepted for publication in a future issue of IEEE Network, but has not been fully edited. Content may change prior to final publication. \textcopyright 2022 IEEE. Personal use of this material is permitted. Permission from IEEE must be obtained for all other uses, in any current or future media, including reprinting/republishing this material for advertising or promotional purposes, creating new collective works, for resale or redistribution to servers or lists, or reuse of any copyrighted component of this work in other works.}
\newcommand\copyrightnotice{%
\begin{tikzpicture}[remember picture,overlay]
\node[anchor=north,yshift=-5pt] at (current page.north) {\fbox{\parbox{\dimexpr\textwidth-\fboxsep-\fboxrule\relax}{\copyrighttext}}};
\end{tikzpicture}%
}

\usepackage{enumitem}
\setlist[enumerate,1]{label={\arabic*.}}

\begin{document}
\addtolength{\textfloatsep}{-2.19pt}

%
\title{Toward Safe and Accelerated Deep Reinforcement Learning for Next-Generation Wireless Networks}

\author{Ahmad M. Nagib,
        Hatem Abou-zeid,
        and Hossam S. Hassanein
        
\thanks{\textit{Ahmad M. Nagib (corresponding author) is with Queen's University, Canada, and also with Cairo University, Egypt. Hatem Abou-zeid is with University of Calgary, Canada. Hossam S. Hassanein is with Queen's University, Canada.} 
        }}


\maketitle
\copyrightnotice

\begin{abstract}
Deep reinforcement learning (DRL) algorithms have recently gained wide attention in the wireless networks domain. They are considered promising approaches for solving dynamic radio resource management (RRM) problems in next-generation networks. Given their capabilities to build an approximate and continuously updated model of the wireless network environments, DRL algorithms can deal with the multifaceted complexity of such environments. Nevertheless, several challenges hinder the practical adoption of DRL in commercial networks. In this article, we first discuss two key practical challenges that are faced but rarely tackled when developing DRL-based RRM solutions. We argue that it is inevitable to address these DRL-related challenges for DRL to find its way to RRM commercial solutions. In particular, we discuss the need to have \textit{safe and accelerated} DRL-based RRM solutions that mitigate the slow convergence and performance instability exhibited by DRL algorithms. We then review and categorize the main approaches used in the RRM domain to develop \textit{safe and accelerated} DRL-based solutions. Finally, a case study is conducted to demonstrate the importance of having \textit{safe and accelerated} DRL-based RRM solutions. We employ multiple variants of transfer learning (TL) techniques to accelerate the convergence of intelligent radio access network (RAN) slicing DRL-based controllers. We also propose a hybrid TL-based approach and sigmoid function-based rewards as examples of safe exploration in DRL-based RAN slicing.

\end{abstract}

\begin{IEEEkeywords}

Next-Generation Networks, B5G, Radio Resource Management, Deep Reinforcement Learning, Safe DRL, Accelerated DRL, RAN Slicing, Transfer Learning

\end{IEEEkeywords}

\IEEEpeerreviewmaketitle

\section{Introduction}

Next-generation networks (NGNs) will support a diverse set of cell and user equipment types, radio access technologies, and communication paradigms. Such multi-level heterogeneity serves a wide range of use cases and deployment scenarios simultaneously. This requires mobile network operators (MNOs) to configure countless network functionalities operating at different timescales and having different objectives \cite{8466370}. Accordingly, the process of optimally configuring such functionalities is not straightforward. A fair amount of these functionalities is linked to efficiently utilizing the limited network radio resources. Hence, this process is called radio resource management (RRM) \cite{9430561}. RRM supports functionalities such as admission control, packet scheduling, and link adaptation. Moreover, it provides functions related to power allocation, beamforming, load balancing, handover management, and inter-cell interference coordination, among others \cite{9372298}.

The complexity of RRM problems is expected to continue growing in NGNs as optimization domains become larger and network requirements become tighter \cite{8466370}. To deal with that, machine learning (ML) techniques have been extensively proposed. Recently, more attention has been given to deep reinforcement learning (DRL) due to its ability to adapt to the dynamic radio access network (RAN) environment in an open control fashion \cite{9372298}. Nevertheless, DRL is known to suffer from convergence issues \cite{9430561}. For instance, it is common for DRL agents to experience a long exploration phase during which drastic performance drops are unavoidable. Exploration refers to trying actions that the agent has not selected before in a given state. By doing so, the agent improves its chances of recognizing the optimal actions \cite{sutton2018reinforcement}.

Exploration is still vital when DRL agents are newly deployed in a live network setting and whenever new contexts or extreme conditions are experienced \cite{9229155, 9457160}. Taking previously unexplored actions might not be optimal for a given condition. This situation might be tolerable in some cases. Nonetheless, in the case of RRM, any potential drops in system performance will affect the end-user's quality of experience (QoE) \cite{9229155}. Such DRL-related practical challenges are rarely tackled in the RRM literature \cite{9430561}. We define \textit{safe and accelerated DRL} as approaches that attempt to avoid system performance instabilities and violations of the network's service level agreements (SLAs). Such techniques also aim at reducing the DRL agents' exploration duration. These approaches are crucially needed to allow the adoption of DRL in commercial NGNs to solve dynamic RRM problems.

The main contributions of this article are summarized as follows:

\begin{itemize}
    \item We discuss and categorize the practical challenges of DRL-based RRM imposed by the exploration phase, and the stochastic nature of the NGNs RAN environments. We make arguments that addressing these challenges is essential for DRL to find its way into commercial solutions.

    \item We present and categorize some of the techniques used to deal with the slow convergence of DRL algorithms and their unstable exploration phase in NGNs. Our proposed categorization gives the readers a structured overview of the approaches that partially tackle such challenges in the context of RRM. It also provides them with insights on choosing a suitable technique.
    
    \item We conduct a case study on intelligent RAN slicing to demonstrate the importance of \textit{safe and accelerated DRL} approaches in NGNs. We highlight the exploration behaviour of various state-of-the-art DRL algorithms. We then analyze the effect of using different reward functions and transfer learning (TL) techniques on the exploration of DRL-based slicing agents. We particularly employ reward shaping, policy reuse, and policy distillation techniques. We finally propose a hybrid technique as an example for safer TL-accelerated exploration. We develop and publicly share the implementation of our OpenAI GYM-compatible DRL environment to enable fellow researchers to further address the discussed challenges.

\end{itemize}

\section{Deep Reinforcement Learning-based Radio Resource Management in NGNs}
\label{RL-RRM}

DRL does not require prior information about the network or access to complete knowledge of the system. Access to such information is inefficient and even inapplicable for the time-varying and uncertain NGNs environments. Hence, the DRL framework is a promising tool to solve the dynamic RRM problem. This capability of DRL has driven use cases related to packet scheduling, power control, handover, and RAN slicing \cite{9372298}.

\begin{figure}
\centering
\includegraphics[width=\linewidth]{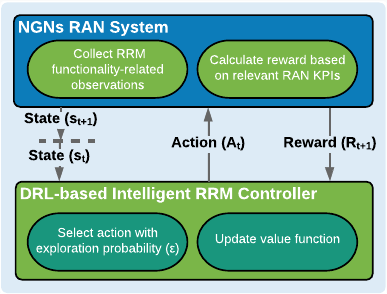}
\caption{The controller–environment interaction in DRL-based RRM in NGNs.}
\label{fig:rl_entities}
\end{figure}

As seen in Fig. \ref{fig:rl_entities}, a DRL-based RRM controller continuously interacts with the RAN environment. At any given time-step, the DRL agent observes the RAN system state and chooses an action to take. Such action changes the RAN environment, and the agent receives reward feedback representing the system's performance. The agent aims at maximizing such feedback. The reward function is designed to guide the agent's search for the optimal policy. It is often represented in terms of a weighted sum of the relevant network's key performance indicators (KPIs). This way the agent indicates how good the action taken was. This is estimated based on the agent's sampled experience from interacting with the RAN environment in a real-time and dynamic-open control fashion \cite{sutton2018reinforcement}.

 \begin{figure*}
\centering
\includegraphics[width=\textwidth]{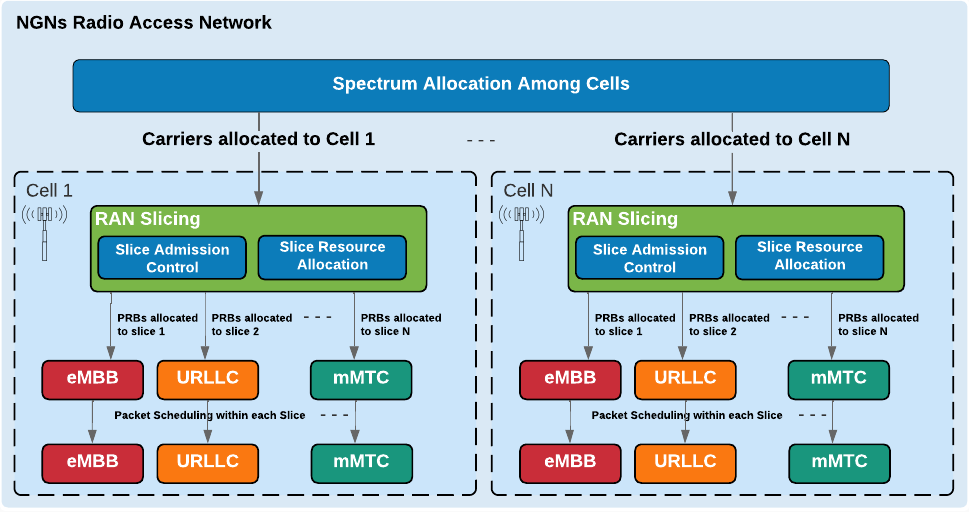}
\caption{Overview of RAN slicing.}
\label{fig:ran_slicing}
\end{figure*}

RAN slicing provides a way to share the physical infrastructure among several services as shown in Fig. \ref{fig:ran_slicing}. It is mainly concerned with two RRM functions. The first is slice admission control in which the infrastructure provider decides whether a service provider's slice request is accepted. This is followed by allocating the available spectrum to the admitted slices. The resources allocated to each slice should enable the slices to comply with their different service requirements. Such constraints should not be violated otherwise monetary penalties can be enforced on the infrastructure provider.

A DRL-based agent is well-suited to this problem \cite{8736403}. The agent builds an approximate and continuously updated model of the RAN's dynamic environment. This way, it can fulfill the different requirements for the various services sharing an infrastructure simultaneously. The agent learns a policy to allocate the available resources to each slice based on the changing network conditions. This includes the channel conditions, the number of admitted slices, the type, and SLAs, of the service supported by each slice, and the number of users in each slice. Moreover, the traffic demand for beyond 5G (B5G) services is dynamic and cannot be easily predicted, particularly in the short term.

\section{Practical Challenges of DRL-based RRM in NGNs}

DRL still has some disadvantages, especially when employed as part of real-time solutions in stochastic environments such as NGNs RAN. We highlight two key DRL-related practical challenges that, even when discussed, are rarely tackled by the wireless networks' community. We encourage researchers to pay more attention to these uncharted territories to speed up the adoption of DRL in RRM commercial solutions.

\subsection{\textbf{Slow Convergence of DRL Algorithms}}

This challenge relates to the number of time-steps it takes the DRL-based controller to find a good set of RRM configurations. The reward feedback that a DRL agent receives may exhibit some variability. Hence, it needs to observe a representative variety of the RAN system's possible states several times. The learning happens by iteratively updating a value function until convergence. This process is called the exploration phase. The value function gives an estimate of the expected return if the agent acts according to a particular policy \cite{sutton2018reinforcement}.

Given the stochasticity of NGNs RAN systems and the exploratory aspect of the DRL agents, it typically takes thousands of time-steps to converge to an optimal configuration for a given RRM functionality. This is of great significance in real network deployments. Only a few exploration iterations can be tolerated in the case of real-time functionalities \cite{9430561}.

\subsection{\textbf{Unstable DRL Exploration Phase}}

RAN systems must maintain a certain performance level to guarantee users' QoE and the overall system's quality of service. Hence, RRM problems are commonly formulated as constrained optimization problems. The DRL agent utilizes the exploration phase to search through the wide spectrum of possible RRM configurations. Hence, there is a high probability that sudden drops in the system performance occur more often. This happens due to exploring RRM configurations that were not encountered previously.

These are not major issues for some applications such as training an agent to play a computer game. Training in this case can be done offline for a long time. Unlike NGNs, the training environment will still match the deployment environment. However, this is a concern in the case of NGNs. The training environment might be simulation-based, and hence, does not precisely reflect the dynamic nature of the real network \cite{9061001}. Thus, in-network learning is needed, but cannot be done for a long time to avoid damaging actions.

\section{Safe and Accelerated DRL-based RRM in NGNs}
\label{approaches}

In this section, we present and categorize research efforts that investigate the aforementioned DRL challenges in the NGNs domain as summarized in Table \ref{tab:Approaches}. We believe that the proposed categorization greatly assists fellow researchers in further addressing the discussed challenges systematically.

\subsection{\textbf{Accelerated DRL-based RRM Solutions}}
The following approaches have shown promise in minimizing the number of learning iterations a DRL agent requires for convergence. 

\subsubsection{\textbf{Domain Knowledge-Aided DRL Acceleration}}

\paragraph{\textbf{Expert Knowledge-Aided Acceleration}}

This category exploits relevant knowledge previously acquired by experts to guide the exploration phase. For instance, the authors of \cite{9847348} propose a structure-aware mechanism to solve a node-overload protection problem in mobile edge computing. In this respect, the optimal policy is known to have a multi-threshold structure. Hence, the agent can reject requests with CPU utilization above a certain threshold. This allows the agent to recover quickly whenever the request distribution changes.

\begin{table*}
\centering

\caption{Practical challenges of DRL-based RRM in NGNs and approaches to tackle them}
\begin{tabular}{|p{1.8in}|p{2.0in}|p{2.4in}|}

\hline
\textbf{Challenge}                                               & \textbf{Approach   category}                              & \textbf{Approach sub-category}                             \\ \hline
\multirow{6}{*}{Slow convergence of DRL algorithms} & \multirow{2}{*}{Domain knowledge-aided DRL acceleration}     & Expert knowledge-aided DRL acceleration \cite{9847348}                    \\ \cline{3-3} 
                                                                        &                                                           & Conventional solution-aided DRL acceleration \cite{8927868}
                      \\ \cline{2-3} 
                                                                        & \multirow{2}{*}{Machine   learning-aided DRL acceleration } & ML-based   experience  building \cite{9229155} \\ \cline{3-3}                      &                                                           & Transfer learning accelerated  DRL \cite{9524965, 9789336}
                                                                    \\ \cline{3-3}                                                                                   &                                                           & Meta-learning    accelerated  DRL \cite{9457160}              
                                                                        \\ \cline{2-3} 
                                                                        & \multirow{2}{*}{Design choices-aided DRL acceleration}     & DRL initialization  strategies \cite{9524965, 5983301}            \\ \cline{3-3} 
                                                                        &                                                           & Inherent  DRL  agent  properties \cite{8736403}                      \\ \hline
\multirow{2}{*}{Unstable exploration phase}                             & \multirow{2}{*}{Safe   DRL-based RRM solutions}            & Transforming the optimization criterion \cite{9259378}         \\ \cline{3-3} 
                                                                        &                                                           & Modifying the exploration process \cite{10.5555/2789272.2886795}                 \\ \hline
\end{tabular}
\label{tab:Approaches}
\end{table*}

\paragraph{\textbf{Conventional Solution-Aided Acceleration}}

Here, a traditional RRM algorithm is used to guide the exploration phase. For instance, in \cite{8927868}, the authors use the proportional fair (PF) algorithm as a separate agent competing with the main DRL agent to solve a resource scheduling problem. The reward is calculated based on the difference in the resulting KPIs between the DRL agent’s action and the PF algorithm. The results suggest that the agent’s performance and convergence speed can be improved.

\subsubsection{\textbf{Machine Learning-Aided DRL Acceleration}}

\paragraph{\textbf{ML-based Experience Building}}

This approach proposes the idea of offline experience building to accelerate training after deploying the agent in a live network setting. The authors of \cite{9229155} proposed this concept in the context of downlink resource allocation for ultra-reliable low-latency communications (URLLC). The experience is built by generative adversarial neural networks that pretrain the DRL agent using a mix of real and synthetic data. This allows the agent to be exposed to a broader range of network conditions. The authors demonstrate that this approach also helps the agent to recover in a few steps whenever it experiences extreme conditions.

\paragraph{\textbf{Transfer Learning Accelerated DRL}}
\label{TL-approaches}

TL expedites learning of new target tasks by exploiting knowledge from related source tasks. This can shorten the learning time of ML algorithms and enhance their robustness to changes in wireless environments. TL techniques have recently emerged as potential solutions to DRL practical challenges such as the long exploration phase in the constantly changing wireless environments \cite{9789336}. In a previous study, we employed a policy transfer approach to accelerate the convergence of DRL-based RAN slicing agents \cite{9524965}. This was based on initializing the policies of newly deployed agents with those of previously trained agents. The results suggest that despite the considerable differences between the traffic models of the source and target scenarios, TL can enhance the convergence behaviour. TL in DRL is further categorized based on the knowledge being transferred, and when and how to transfer such knowledge. We demonstrate four approaches belonging to two of these sub-categories in the case study section.

\paragraph{\textbf{Meta-learning Accelerated DRL}}

Meta-learning was introduced in the context of supervised ML to design models that can learn new skills or adapt to new environments with a few training examples. The same concept can be used to accelerate DRL agents' convergence as in \cite{9457160}. The authors developed a DRL-based solution to control drone base stations (BSs) providing uplink connectivity to ground users. The trained policy should satisfy the users' dynamic and unpredictable access requests. Meta-training is employed to find, for every user request realization, a set of initial policy and value functions that are close to the optimal ones. This can be fulfilled by minimizing the losses that are collected from the sampled user request realizations. This is done while serving the ground users' daily requests as an attempt to generalize the learning to unseen environments. Thus, unlike TL, this does not require prior knowledge from agents previously trained on similar tasks. The authors demonstrate that the meta-trained agent yields faster convergence to the optimal coverage in unseen environments.

\begin{table*}
\centering
\caption{Simulation parameters and DRL design details}
\begin{tabular}{|p{1.3in}|p{1.4in}|p{1.4in}|p{1.6in}|}
\hline
\multicolumn{4}{|c|}{\textbf{(a) RAN slicing simulation parameter settings}}                                                                                                                                                                                                                                                                                                            \\ \hline
                                                                                                       & \textbf{Video}                                     & \textbf{VoLTE}                                                                        & \textbf{URLLC}                                                                                                                    \\ \hline
\textbf{Scheduling algorithm}                                                                          & \multicolumn{3}{l|}{Round-robin per 0.5 ms slot}                                                                                                                                                                                                                               \\ \hline
\textbf{Bandwidth allocation window size}                                                              & \multicolumn{3}{l|}{40 scheduling time slots}                                                                                                                                                                                                                     \\ \hline
\textbf{\begin{tabular}[c]{@{}l@{}}Packet interarrival time \\ distribution \end{tabular}}   & Truncated Pareto (mean = 6 ms, max = 12.5 ms)      & Uniform (min = 0 ms, max = 160 ms)                                                       & Exponential   (mean = 180 ms)                                                                                                     \\ \hline

\textbf{\begin{tabular}[c]{@{}l@{}}Packet size distribution\\ \end{tabular}}                & Truncated Pareto (mean = 100 B, max = 250 B) & Constant (40 B)                                                                    & Truncated log-normal (mean = 2 MB, standard deviation = 0.722 MB, max = 5 MB)                                                      \\ \hline
\multicolumn{4}{|c|}{\textbf{(b) RAN slicing DRL design}}                                                                                                                                                                                                                                                                                                                               \\ \hline
\multicolumn{2}{|l|}{\textbf{State}}                                                                                                                        & \multicolumn{2}{l|}{\begin{tabular}[c]{@{}l@{}}The ratio of slices' traffic load in the last slicing window \\ $(p_{Video}, p_{VoLTE}, p_{URLLC})$\end{tabular}}                                                     \\ \hline
\multicolumn{2}{|l|}{\textbf{Action}}                                                                                                                       & \multicolumn{2}{l|}{\begin{tabular}[c]{@{}l@{}}The percentage of bandwidth allocated to each slice (15 allocation configurations)\\ $(w_{Video}, w_{VoLTE}, w_{URLLC})$, s.t. $w_{Video} + w_{VoLTE} + w_{URLLC} = 100\%$\end{tabular}} \\ \hline

\multicolumn{2}{|l|}{\textbf{Reward function 1}}                                                                                                                       & \multicolumn{2}{l|}{A weighted sum of the slices' average latency in a slicing window}                                                                                                                           \\ \hline
\multicolumn{2}{|l|}{\textbf{Reward function 2}}                                                                                                                     & \multicolumn{2}{l|}{A weighted sum of a sigmoid
function-based reward with slices' latency as a variable}                                                                                                                           \\ \hline
\multicolumn{2}{|l|}{\textbf{Reward function 3 (for reward shaping)}}                                                                                                                    & \multicolumn{2}{l|}{Same as function 2 with extra reward when URLLC slice requirements are satisfied}                                                                                                                     \\ \hline
\multicolumn{2}{|l|}{\textbf{Reward function slice weights}}                                                                                                                       & \multicolumn{2}{l|}{URLLC: 0.7, Video: 0.2, VoLTE: 0.1}                                                                                                                         \\ \hline
\multicolumn{2}{|l|}{\textbf{Hard slicing resource allocation}}                                                                                                                       & \multicolumn{2}{l|}{URLLC: 33\%, Video: 33\%, VoLTE: 33\%}                                                                                                                       \\ \hline
\multicolumn{2}{|l|}{\textbf{Fixed slicing resource allocation}}                                                                                                                     & \multicolumn{2}{l|}{URLLC: 70\%, Video: 20\%, VoLTE: 10\%}                                                                                                                       \\ \hline
\multirow{5}{*}{\textbf{DRL parameters}}                                                               & \textbf{DRL algorithms}                            & \multicolumn{2}{l|}{Dueling DQN, PPO, A2C, REINFORCE, TRPO + hard slicing and fixed slicing    }                                                                                                                             \\ \cline{2-4} 
                                                                                                       & \textbf{Experiment time-steps}                & \multicolumn{2}{l|}{Expert BS: 50,000, learner BS: 20,000}                                                                                                                                                             \\ \cline{2-4} 
                                                                                                       & \textbf{Exploration}                               & \multicolumn{2}{l|}{Expert BS: 0.9, learner BS: 0.2  }                                                                                                                                                       \\ \cline{2-4} 
                                                                                                       & \textbf{Exploration decay}                         & \multicolumn{2}{l|}{Expert and learner BS: 0.99 }                                                                                                                                                       \\ \cline{2-4} 
                                                                                                       & \textbf{Batch size}                                & \multicolumn{2}{l|}{4-8}                                                                                                                                                                                    \\ \hline
\end{tabular}
\label{tab:sim_parameters}
\end{table*}

\subsubsection{\textbf{Design Choices-Aided DRL Acceleration}}
\label{design}

Unlike the previously discussed methodologies, the following approaches rely on intrinsic DRL properties. The idea is to make efficient DRL design choices to shorten the exploration duration.

\paragraph{\textbf{DRL Initialization Strategies}}
\label{init}
Several parameters can be specified for a DRL agent. This includes the initial policy, learning rates, and neural network architecture. The authors in \cite{5983301} propose a decentralized approach for interference management between femtocells and macrocells. To overcome the slow convergence of the Q-learning algorithm, they propose a Q-table initialization procedure. Given a new state, the Q-value of the action taken is updated, and the costs of the other actions are estimated.

\paragraph{\textbf{Inherent DRL Agent Properties}}

Some wireless network studies rely on the inherent properties of the DRL algorithms when choosing an algorithm to employ. This includes picking a DRL algorithm and a hyper-parameter setting over another based on their relative performance in terms of stability and convergence time. Authors of \cite{8736403} propose a collaborative learning framework for resource scheduling in RAN slicing. The authors chose the asynchronous advantage actor-critic (A3C) algorithm as it is known to have faster convergence compared with the actor-critic (AC) algorithm.

\subsubsection{\textbf{Choosing an Acceleration Strategy}}
The knowledge available at the time of training influences the choice of the acceleration strategy. For instance, a domain knowledge-aided approach can be employed if relevant prior knowledge is available or a traditional RRM strategy is known. Moreover, ML-based experience building can be used if network data is available and a smooth transition between offline simulation and live network deployment is required. Meta-learning can be used without prior knowledge about the task. It considers generalization while learning so that an agent starts with a near-optimal policy in new situations. TL requires knowledge from agents previously trained on similar tasks. This can be, among others, saved policies, or reward functions. The design choices-aided acceleration should be considered regardless of the available knowledge form. Finally, some approaches are restricted to specific DRL algorithms so this should also be considered while choosing a strategy \cite{9789336}.

\subsection{\textbf{Safe DRL-based RRM Solutions}}
\label{safe-drl-solutions}
It is essential to have a means of safe exploration in deployed DRL-based RRM solutions. The approaches that attempt to speed up the DRL convergence do not necessarily guarantee the avoidance of large negative outcomes, particularly in stochastic wireless environments. However, despite not being designed to address safety, mitigating slow convergence would indirectly reduce the times an agent performs damaging actions.

The unstable exploration phase challenge motivates the development of DRL algorithms explicitly designed to provide safety measures. Safe DRL is the process of learning policies that maximize the value function when some safety constraints must be imposed. This is not restricted to the performance during the learning process but also after deployment \cite{10.5555/2789272.2886795}. The safety concept does not necessarily refer to physical safety (such as in robotics environments). It can also be extended to situations when an agent following a specific policy performs poorly in dynamic environments. Such poor performance may still happen while the agent follows an optimal policy, since maximizing the long-term reward does not necessarily avoid the rare occurrences of undesired outcomes.

Safe DRL can be classified into two main categories \cite{10.5555/2789272.2886795}: 
\begin{enumerate}
    \item Transforming the optimization criterion to include a form of risk.
    \item Modifying the exploration process. 
\end{enumerate}

The second category can be fulfilled either by external knowledge or risk metric incorporation. Constrained DRL is among the main approaches that fall into the first category. It is an extension of the DRL framework where a set of constraints applied to the policy are integrated. This can help RRM by enforcing bounds to guarantee that the DRL-based RRM controller will not violate any constraints posed by the MNOs. In other words, the controller will be obliged to avoid choosing RRM configurations that might lead to unacceptably low-performance levels. Authors of \cite{9259378} use this approach to enforce both cumulative and instantaneous constraints on network slicing resource allocation. This allows the RAN system to stick to the requirements of the services supported by the various RAN slices. 

Multiple criteria should be analyzed when choosing a safe DRL approach. For instance, the worst-case criterion, which falls under the first category of safe DRL, is useful when rare occurrences of large negative returns must be avoided. The reader is referred to \cite{10.5555/2789272.2886795} for a comprehensive analysis of such criteria. Only a few studies in the NGNs domain explicitly address the challenge of unstable DRL exploration. Hence, more effort needs to be directed toward this area of research given the uncertainty of RAN environments. With critical B5G applications and industrial automation, downgraded performance will not be tolerated since this will result in unacceptable reliability measures for these services.

\section{Intelligent RAN Slicing: A Case Study}
\label{case-study}

We demonstrate the need, and impact of various approaches for \textit{safe and accelerated DRL} in the resource allocation problem of RAN slicing.

\subsection{\textbf{Experiment Setup}}
\label{experiment_settings}

Slicing can be done on many levels. In our experiment, we focus on the BS level. The environment state reflects the traffic load for each slice relative to the total BS load within a previous time window. The DRL agent takes an action to allocate the available PRBs to the admitted slices. Then, round-robin scheduling is carried out independently within each slice. Scheduling multiple transmissions per TTI is supported if resources are available. Requests are generated based on the parameters described in Table \ref{tab:sim_parameters}. Unsatisfied users with multiple unfulfilled transmission requests leave the system. 

It is important to design a reward function that reflects the requirements of the various slices. Here, we focus on the delay requirements. Latency is relatively more important in URLLC slices. Hence, a large weight is configured for the URLLC slices in the weighted sum reward functions as seen in the table. Reward function 1 reflects the weighted sum of the slices’ average latency in a slicing window.

We first evaluate various state-of-the-art DRL algorithms implemented in the Tensorforce Python package to show the exploration performance when using reward function 1. We use a hard slicing agent as a baseline that assigns the spectrum equally among the available slices. Moreover, fixed slicing statically allocates the PRBs according to the percentages in Table \ref{tab:sim_parameters}.

\subsection{\textbf{Methods for Safe and Accelerated DRL-based RAN Slicing}}
\label{TL-demo}

We demonstrate how safety and acceleration can be achieved in DRL-based RAN slicing using the following approaches. Given the scenario, the proximal policy optimization (PPO) agent experienced relatively more frequent drops in rewards as highlighted in the results section. Thus, we decided to employ PPO as the underlying algorithm in all the approaches. However, the other algorithms showed similar behaviour. We use a time-decayed exploration parameter, $\epsilon$, to control the exploration-exploitation trade-off \cite{sutton2018reinforcement}. We start with inspecting the effect of using different reward functions on exploration behaviour. We then demonstrate three approaches that belong to two sub-categories of TL in \cite{9789336}, namely, reward shaping and policy transfer. Furthermore, we propose a fourth approach that falls under the policy transfer sub-category.

\subsubsection{\textbf{Reward Function Design}}
\label{function_design}

\begin{figure}
\includegraphics[width=1\linewidth,left]{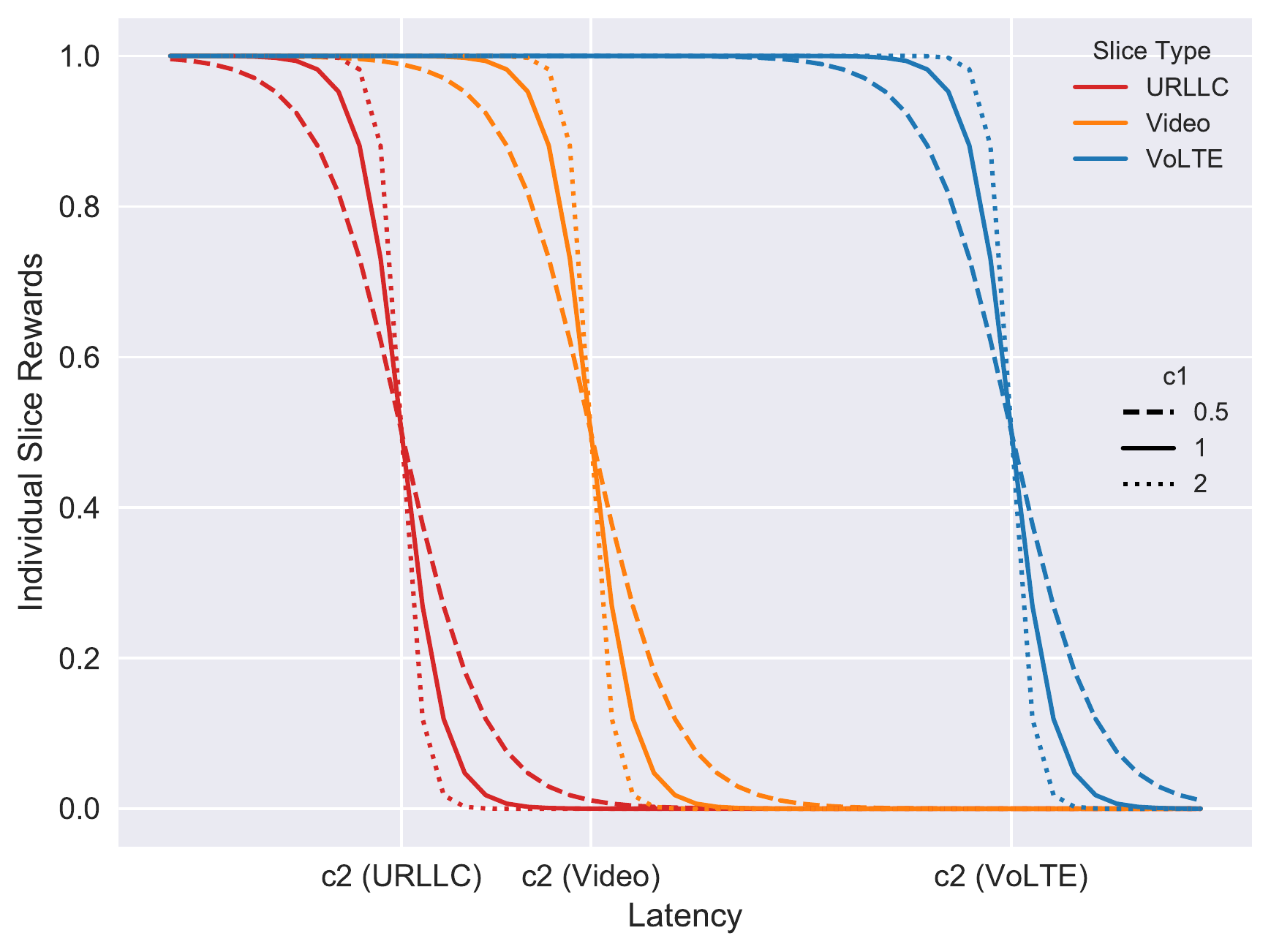}
\caption{An example of reward function 2: $c1$ decides the point to start penalizing the agent's actions; and $c2$ reflects the acceptable latency for each slice.}
\label{fig:sigmoid_function}
\end{figure}

\paragraph{\textbf{Reward Function Change}} The first designed function is a weighted sum of the slices' average latency in a slicing window. We additionally design a second sigmoid function-based reward with latency as a variable. Unlike the first function, function 2 penalizes actions that come close to violating slices' latency requirements. Two parameters, $c1$ and $c2$, are configured to tune the shape of the function as seen in Fig. \ref{fig:sigmoid_function}. $c2$ reflects the minimum acceptable latency for each slice, while $c1$ determines when to start penalizing the agent's actions. The second function follows a safe DRL approach similar to the risk-sensitive criterion in \cite{10.5555/2789272.2886795}. It is a subcategory of transforming the optimization criterion mentioned earlier, in which a parameter is used to enable the sensitivity to the risk to be controlled.

\paragraph{\textbf{Reward Shaping}} Reward shaping uses external knowledge to render auxiliary rewards that guide the DRL agent toward the desired policy. This can help the agent reach an optimal policy faster. We demonstrate this by defining reward function 3 where additional rewards are provided whenever an action leads to satisfying the URLLC latency requirements. As previously mentioned, we prioritize URLLC slices due to their intolerance to delay.

\begin{figure*}[ht]
     \centering
     \begin{subfigure}[b]{0.49\linewidth}
         \centering
         \includegraphics[width=1\linewidth]{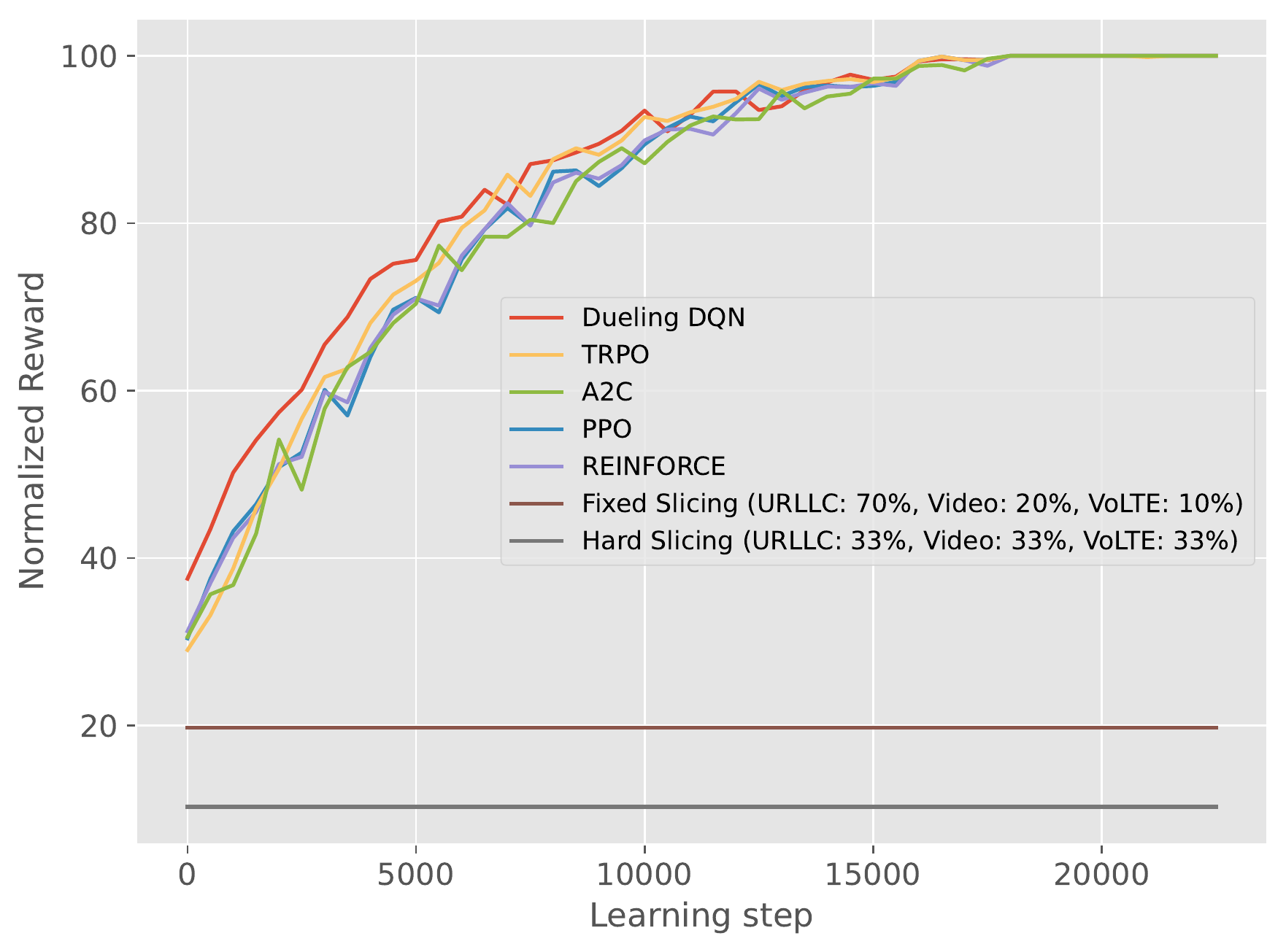}
         \caption{}
         \label{fig:reward_comparison}
     \end{subfigure}
     \hfill
     \begin{subfigure}[b]{0.49\linewidth}
         \centering
         \includegraphics[width=1\linewidth]{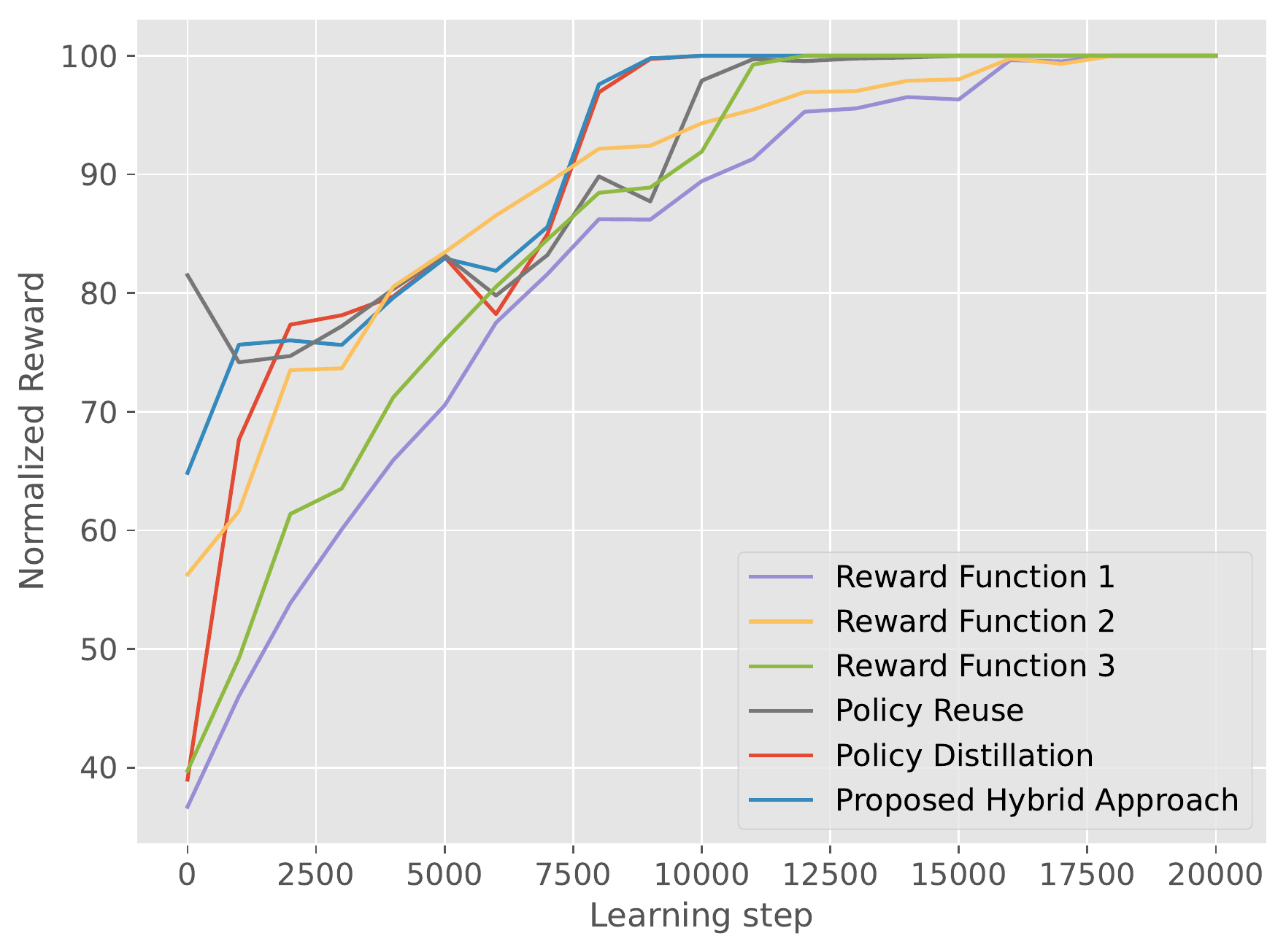}
         \caption{}
         \label{fig:accelerated_reward_comparison}
     \end{subfigure}
        \caption{Numerical results: a) DRL-based RAN slicing exploration performance using various algorithms. Reward is smoothed using a moving average of window size = 500 time-steps; b) comparison of acceleration and safety approaches for DRL-based RAN slicing. Reward is smoothed using a moving average of window size = 1000 time-steps.}
        \label{fig:numerical_results}
\end{figure*}

\subsubsection{\textbf{Policy Transfer}}

Policy transfer is a class of TL in which a source policy is transferred to an agent with a similar target task \cite{9789336}. We employ policy transfer in the following three approaches. We first train an agent to learn a policy at a BS, namely, the expert BS. This scenario includes three slices reflecting the URLLC, Video, and VoLTE services. The expert policy is then used to guide newly deployed agents at other BSs, namely, learner BSs to accelerate their learning processes. The learner BSs scenario includes one URLLC slice and two VoLTE slices.

\paragraph{\textbf{Policy Reuse}}

Here, a source policy that is learned at an expert BS is directly reused to guide the target policy at a learner BS \cite{9789336}. We configure the learner BS to follow the expert policy for the first 500 time-steps. The target policy is continuously updated based on the reward feedback the learner agent receives. We use a transfer rate, $\theta = 1$, during these time-steps meaning that we always follow the source policy. We follow the target policy afterward. However, a smaller or different decaying transfer rate can be configured to switch between the source and target policies during exploration \cite{9789336}.

\paragraph{\textbf{Policy Distillation}}

Here, one or more source policies are combined to guide an agent in a similar target task \cite{9789336}. This can be done by minimizing the divergence of action distributions between the source and the target policies. We follow a similar training approach by minimizing the Euclidean distance between the actions recommended by the expert policy and those recommended by the learner policy. The learner BS is configured to follow the distilled policy for the first 1000 time-steps during which a transfer rate, $\theta = 1$ is used.

\paragraph{\textbf{Hybrid Policy Reuse and Distillation}}
We propose a hybrid of the last two approaches to achieve a safer TL-accelerated exploration. This is helpful when the transferred policies are not generic enough to robustly adapt to new traffic patterns. We introduce a parameter similar to $\theta$ to balance between exploiting the expert policy and exploring a distilled action. This is done during the first 700 time-steps while the target policy is continuously updated based on the reward feedback the learner agent receives. The updated target policy is then followed for the rest of the training phase.

The simulation environment where all these methods are implemented is available on GitHub\footnote{\href{https://github.com/ahmadnagib/SARL-RRM}{http://www.github.com/ahmadnagib/SARL-RRM}}. This allows researchers to study and easily compare the behaviour of the developed DRL-based RAN slicing controllers. The developed environment follows the standard OpenAI GYM interface\footnote{\href{http://gym.openai.com/}{OpenAI Gym: http://gym.openai.com/}}. This enables researchers to develop algorithms that work instantly without any changes to the environment. It also gives researchers a wide spectrum of GYM-compatible software libraries to choose from.

\subsection{\textbf{Numerical Results and Discussion}}
\label{numerical_results}

We first illustrate the exploration performance of the various DRL algorithms in Fig. \ref{fig:reward_comparison}. Hard, and fixed slicing have the lowest reward values as they do not explicitly consider the latency. The results highlight the challenge of slow convergence of DRL algorithms. It can take an agent more than 17,000 learning steps to converge. This is a concern in NGNs because live networks cannot tolerate a non-optimal performance for a long duration. Exploration is needed primarily in two situations: 
\begin{enumerate}
  \item When an agent is newly deployed in a live network. This would happen even for DRL agents trained in simulation-based environments. Such environments fail to accurately reflect the multi-faceted complexities of the dynamic NGNs across all deployments \cite{9229155}.
  \item Whenever the network context changes significantly. This situation is typical of highly dynamic environments, such as when DRL-based drone BSs need to provide coverage to ground users in previously unseen environments \cite{9457160}.
 \end{enumerate}

In Fig. \ref{fig:accelerated_reward_comparison}, we show the results of the approaches described in this case study. We observe the following:

\begin{itemize}
  \item The agent using reward function 1 has the lowest reward throughout most of the exploration phase. This is because function 1 reflects the average latency in a slice without considering the latency requirements for each service type.

  \item Reward function 2 explicitly includes the latency requirements and penalizes the agent whenever it takes actions that violate them. Hence, the agent using function 2 has an enhanced exploration performance. However, it still converges close to the 20,000 steps mark.

  \item Reward shaping succeeds in accelerating convergence. The auxiliary rewards consistently guide the agent toward satisfying the latency requirements of URLLC.

  \item The two policy transfer approaches performed relatively better throughout most of the simulation in terms of convergence time and reward value. However, they have considerably different behaviour at the beginning of the simulation. This is mainly a result of the nature of each approach.

  \item Policy reuse starts with a relatively high reward value as the transferred policy recommends actions close to the optimal ones. On the other hand, policy distillation starts with a lower value as it tries to reduce the divergence between the actions recommended by the learner policy and the expert policy. This smooths out the reward to some value in between. However, it converges faster as it explores more actions at the beginning rather than steadily following a transferred policy.

  \item Both policy transfer techniques experience significant performance drops, unlike the previous approaches. This is mainly because they both rely on a non-generic expert policy. This affects exploration robustness and hence the end-users' QoE.

  \item The proposed hybrid approach combines a good starting reward value, a more stable exploration performance, and a fast convergence rate. This is mainly because it strikes a balance between relying totally on an expert policy trained on a specific scenario and learning from scratch.
\end{itemize}

\section{Conclusion}

The deployment of DRL-based RRM solutions in real networks is subject to several challenges given the uncertainty of NGNs' RAN environments. \textit{Safe and accelerated} exploration is an essential concept that will open the door to DRL-based RRM commercial solutions. Our case study on intelligent RAN slicing demonstrates that DRL agents can take thousands of learning time-steps to converge to a good policy. This results in violations of the various slices' SLAs, and consequently, monetary penalties and undesirable QoE. Our experiment highlights the potential of using transfer learning to guide the exploration process. Moreover, we propose a hybrid approach as an example of safe TL-accelerated exploration. Utilizing acceleration strategies does not guarantee a certain desired performance level. Hence, more effort should be directed toward innovating safe techniques that guarantee the instantaneous and cumulative RRM constraints in NGNs.

\bibliographystyle{IEEEtran}

\bibliography{references}

\section*{Biographies}
\textsc{Ahmad M. Nagib} [GS] (ahmad@cs.queensu.ca) is a Ph.D. student and graduate research fellow at the School of Computing, Queen’s University. He received his B.Sc. and M.Sc. degrees from the Faculty of Computers and Artificial Intelligence, Cairo University. He also works there as an Assistant Lecturer. He is currently part of an industry-academia collaboration project with Ericsson, Canada. His research mainly addresses the practical challenges of applying machine learning, and specifically reinforcement learning, in next-generation wireless networks. He served as a TPC member and reviewer in several IEEE flagship venues such as TNSM, GLOBECOM, ICC, and LCN.

\hfill

\textsc{Hatem Abou-Zeid} [M] (hatem.abouzeid@ucalgary.ca) is an Assistant Professor at the University of Calgary. Prior to that he was at Ericsson leading 5G radio access research and IP in RAN intelligence, low-latency communications, and spectrum sharing. Several wireless access and traffic engineering techniques that he co-invented and co-developed are deployed in mobile networks and data centers worldwide. His research interests are broadly in 5G/6G networks, extended reality communications, and robust machine learning. His work has resulted in 19 patent filings and 50 journal and conference publications in several IEEE flagship venues. He received the PhD degree from Queen’s University in 2014.

\hfill

\textsc{Hossam S. Hassanein} [S’86, M’90, SM’05, F’17] (hossam@cs.queensu.ca) is a leading authority in the areas of broadband, wireless and mobile networks architecture, protocols, control and performance evaluation. His record spans more than 600 publications in journals, conferences and book chapters, in addition to numerous keynotes and plenary talks in flagship venues. He has received several recognition and best paper awards at top international conferences. He is a Fellow of the IEEE and is a former chair of the IEEE Communication Society Technical Committee on IoT, AdHoc and Sensor Networks. He is an IEEE Communications Society Distinguished Speaker (Distinguished Lecturer 2008-2010).

\end{document}